# Environmental and Economic Impact of I/O Device Obsolescence

Patrick Gould
College of Engineering
*The Ohio State University*
Columbus, OH, USA
gould.279@osu.edu

Guanqun Song
College of Engineering
*The Ohio State University*
Columbus, OH, USA
song.2107@osu.edu

Ting Zhu
College of Engineering
*The Ohio State University*
Columbus, OH, USA
zhu.3445@osu.edu

*Abstract*—This paper analyzes the proportion of Input/output devices made obsolete by changes in technology generations. This obsolescence may be by new software/hardware generations rendering otherwise functional devices unusable. Concluding with brief analysis on the economic and environmental impacts of the e-waste produced.

*Keywords— Operating systems, Obsolescence, Environmental, Economic, I/O*

## I. INTRODUCTION

As the world becomes ever more reliant on digital technology, technological innovation grows with it. However, as the world continues to adopt new technologies and devices in various fields like IoT [38-55], machine learning [60-68], security field[56-59], etc., old devices are abandoned as they are made obsolete by being replaced by better alternatives, or, by being made useless by design of its creators. This practice, called 'forced obsolescence' has existed for many years now [1]. Forced obsolescence by means of restricting driver support, changing ports, creating devices with unreplaceable disposable batteries. These are some of the more common ways of manufacturers try to pressure consumers to buy new devices and discard older, otherwise functional, ones. The result of this is more money spent by the consumer and an ever-accelerating amount electronic waste (henceforth called e-waste) dumped back into the environment. Much of this e-waste is not collected or properly recycled. As the world's ecosystem becomes ever more unstable, the importance of drawing attention to reducing waste that is heavy in metals, plastic, and other environmentally expensive materials has become important. There is great deal of work looking at the wide-ranging effects of e-waste on both the economy and the environment, however, it is often agnostic to why the e-waste is produced. That is, there is little research looking into the important role operating systems play and how we might better design them to consider things like their economic and environmental effects. In this paper, we are going to analyze: The devices that are interoperable between major OS's, the proportion of devices made obsolete by new software generations, proportion of e-waste collected to be recycled, proportion of e-waste that is able to be recycled, and consider the environmental and economic impacts of all of these. To prevent the focus of this paper from becoming too broad, it shall focus primarily on input/output devices (henceforth called I/O devices, I/O).

## II. RELATED WORK

There is a great deal of work and research done on the topic of e-waste, ways to better recycle it, and ways to better reduce it. Much of the work relating to e-waste is broadly done by governmental agencies, colleges and universities, and environmental or political organizations.

### A. Governmental

Governments put forth a great deal of effort in finding ways to reduce or recycle various forms of electronic waste—much of which does not cleanly biodegrade like other waste. Much of the work takes the form of large-scale statistical studies. The World Health Organization [2] published a quantitative review of the transboundary "flows" of used electronics. The methods in the report involved gathering of large sets of public information for common devices that become e-waste and preforming the following:

1. Determining the sales of a product in a region over a time period: This involves gathering historical sales data, which provides a baseline for estimating how many devices are likely to enter the waste stream in the future.

2. Determining the typical distribution of lifespans for the product over a time period: By analyzing how long products are typically used before being discarded, researchers can better predict the timing of e-waste generation.

3. Calculating how many products are predicted to be generated in a given year: Using the sales and lifespan data, the total volume of e-waste generated annually can be estimated.

4. Calculating how many of the generated products are predicted to be collected in a given year: By applying regional collection rates, governments can estimate how much e-waste is successfully collected for recycling.

5. (Optional) Calculating the weight of generated and collected products: This involves multiplying unit weights by the quantities estimated, which helps assess the physical and logistical challenges of managing e-waste.

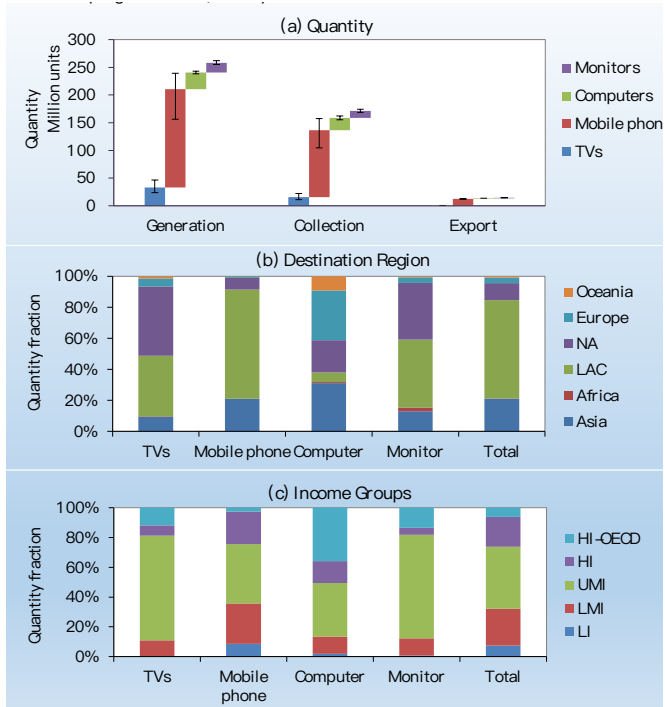

*Figure 1 Proportion of device collected in the United States*

Note the large difference in the generation and collection rates, this will be interesting later. The paper concludes with the recommendation that for future research to be conducted there is a need for "The creation of trade codes for used products would enable explicit tracking of those products." An important thing to note from the study is that the world does not really have all the information its needs to be able to properly study and solve the issues around e-waste. Another common form of work done is in the interest of public and environmental health: the Environmental Protection agency (EPA) [3] funded and put forth a report on the challenges in tracking information relating to e-waste. This resulted in the formation of several different studies in cooperation with the World Health Organization.

*B. Academic*

Michigan University [4] has done work to make "informal" recycling of materials safer by developing inexpensive tools and protective equipment for workers. As an extra side effect, help keep hazardous materials out of landfills and local environments. Additionally, since some of this work is often done in homes, proper tools and equipment also keeps living areas, working space, and food preparation safer.

*C. Organizational*

Other work is done by organizations, often through public education, and, quite often, passing the previous governmental and academic sources to the public in more digestible format.

### III. DESIGN

The majority of the content of this paper is concerned with rather wide-ranging effects of e-waste and why it is made. So, the design of experimentation is consider large, existing studies and analyzing the results, with the intent to answer the following questions:

1. What I/O devices (external monitors, keyboard, trackpad, earplug, docking stations, etc. ) are only workable for Mac? What accessories are only workable for Windows?
2. With the new release of Mac or windows computers, how many of the above accessories become obsolete?
3. How will these impact the environment? How to reuse, recycle, and disposal of these old accessories?
4. What's the economic impact?

### IV. Evaluation

To begin: "What I/O devices (external monitors, trackpad, earplug, docking stations, etc. ) are only workable for Mac? What accessories are only workable for Windows?" Fortunately, Apple provides [5] a clear set of accessory (I/O) guidelines for Apple devices, including what standards a given device must to comply with. Unfortunately, some classes of devices are missing from this specification, such as monitors or docks. What (nonproprietary) device classes are given are: Port/cable adaptors, AC power adaptors, battery packs, headsets, strobes (camera flash) keyboards, trackpads, external storage.

*A. HID devices (Keyboards, mice, game controllers, etc.)*

Many devices noted by Apple fall under the umbrella of 'Human Interface Device' (HID)—anything that a person uses to directly interact with a given computer; anything with a button, effectively. HID is a standard defined by the USB Implementers Forum [6] and is also implemented by Bluetooth [7]. Apple specifies the following for HID devices:

- If connected over USB: "If implementing HID over USB, the accessory shall comply with the Device Class Definition for Human Interface Devices 1.11 [8].
- If connected over Bluetooth: Should support Bluetooth HID Profile 1.1 [7].

Conveniently, Windows specifies [9] these exact same standards for all HID devices. That is, if a HID device works on one operating system, it must work on the other—assuming a given computer has proper hardware to support a device. E.g., most PCs do not support Xbox wireless controllers without a proprietary adaptor [10], but Macs that can support macOS "Catalina" [10] can. This guarantee does not apply to extra 'specialty' hardware features of I/O devices, like Apple's TouchID sensor on their Magic Keyboard, may not work on other operating systems without special made drivers. Fortunately, for the devices that implement the HID standards as defined in [7][8], they will otherwise function normally. I.e., you may connect and use a TouchID enabled Magic keyboard on a PC, you simply will not be able to use TouchID function. A similar case may be found for Mac: Razer, a gaming hardware company, manufactures a mouse called "Mamba

Wireless" that has several special settings, like setting RGB light values, that may only be adjusted using proprietary software. This software, called "synapse" is not supported on Macs using macOS versions newer than 10.10 [11]. Although, from testing, functions correctly as a pointing device otherwise. The precise proportion of these devices on the market with such restrictions, but are otherwise functional, is not well understood and is suggested as a topic for further research.

*B. Monitors*

Monitors are not noted in any Apple documentation pertaining to developers. That is, exact standards are not noted. Although, all modern Macs (released in the last 10 years) are USB-C, DisplayPort, or HDMI compliant [12], which are common standards for displaying both video and audio. Windows, by contrast, gives rather extensive [13] guidelines for all monitors, although only notes a single standard for connection, DisplayPort—although similar standards used by Apple are also used in practice. I.e., HDMI, USB-c Thunderbolt. In writing this report, no monitors were found to explicitly exclude a certain operating system but work on others; both Windows and macOS use common standards. The only notable exceptions are monitors that Apple manufactures which require thunderbolt 3.0 [14][15], a standard uncommon for many computers. This appears to be due to the fact that Thunderbolt may use up to four PCI lanes [16]. This is convenient for mobile devices, like laptops, which may not have room for devices connected using more traditional PCI slots. Desktops, by contrast, do not have such a restriction, so thunderbolt connections using PCI lanes is a less reasonable trade-off. Additionally, monitors made by Apple have some proprietary software that does not work on PCs [17].

*C. Earplugs*

Headphones, and other audio device, are a common exemplar class of I/O devices. Similar to HID devices, audio transferred over common standards like Bluetooth, USB, or 3.5 mm headphone jack (although becoming uncommon on Apple devices) will work across operating systems. AirPods Pro, for example, are earbuds made by Apple that transmit audio over Bluetooth and work on all Bluetooth enabled Apple devices. As a test of this, a pair of these headphones were tested on a PC using Windows 11:

- AirPods Pro Model: A2084

- Firmware version: 6A321 (latest version at time of writing)

- PC operating system: Windows 11 (OS Build: 22631.3447)

Speakers and microphone worked as expected. Additionally, noise canceling and pressure sensor (acts as a button) worked correctly as well. This is to say, that unless an audio is using an unusual transmission standard or specialty software, it will be interoperable between major operating systems. Similar to HID devices, the exact proportion of devices that have features that do not work between operating systems is not well known.

*D. Other devices*

Other common classes of devices, like USB docks, are similar. If a device uses the aforementioned connection standards, they should be interoperable between operating systems. There is a notable category of external GPUs, that are connected through a thunderbolt connection, that belong here and are pertinent to the next discussion.

*E. Case Study: Blackmagic eGPUs*

An interesting category of I/O devices are external GPUs (called eGPU henceforth). These processors are used, as the name suggests, for graphical processing for systems without strong GPUs. Blackmagic, specifically, is a brand of eGPU that was officially supported by Apple, they even sold them on their store [18]. However, with the release of Apple's new class of processors, M1 [19], it's new computers no longer supported any kind of eGPU. This is not because of an operating restriction but instead an architectural one, macOS still supports eGPUs on Intel processors, just not Apple ones [20]. All of this does show that designers of major operating systems may create, or possibly destroy, entire classes of devices within a hardware generation. The exact number of eGPUs made obsolete is not known, Apple nor Blackmagic make relevant sales numbers public at time of writing.

*F. Case Study: Windows 11*

A related issue is with Windows 11. When Microsoft released Windows 11 they raised the required hardware specifications [21], rendering roughly 240 million [22] devices no longer able to receive normal updates by October 14, 2025 [23]. The precise amount of waste produce from this is not well known, many media sources claim that these computers comprise roughly "480 million kilograms" of potential waste, first appearing in a publication from Reuters [24], but notably, this number is never justified.

*G. Power adaptors*

It is notable that AC power blocks and charging cables are not listed above. Broadly, this is because charging I/O devices are often hardware specific and not operating systems specific. However, many devices require a proprietary charging cables or blocks. These chargers cannot always be used between systems. A common example is Apple's Lightning cables—which are defined in its style guide [5]. An interesting development to this topic is that the European Union (EU) now requires [25] all new mobile devices sold in the EU must have a standardized USB type-c charging port. Mobile devices are defined as:

- mobile phones

- tablets

- digital cameras

- headphones

- headsets

- handheld videogame consoles

- portable speakers

- e-readers
- keyboards
- mice
- portable navigation systems
- earbuds
- laptops

While this legislation does not necessary affect other regions, having such a large proportion of the global market use a single standard does create a strong incentive for manufacturers to standardize their I/O across borders. It is also worth noting that there will likely be a 'one time' dump of e-waste as new standards take place and old I/O devices are replaced.

An interesting conclusion to draw from each of these considerations above is that for most traditional I/O devices, like mice, keyboards, or monitors, they are almost always interoperable between systems, excusing any hardware restrictions.

"With the new release of Mac or windows computers, how many of the above accessories become obsolete?" When considering USB, and other common I/O devices, it appears to be very little. The USB 4.0 specification [26], the latest version, requires a minimum backwards compatibility with USB 2.0—released in 2000 [27]. Similarly, HDMI [28], DisplayPort [29], and Bluetooth [30] have so far been backwards compatible with all previous versions. This points to a rather interesting trend in I/O devices: an emphasis on backwards compatibility and longevity. So, the proportion of I/O devices that are rendered obsolete with successive computer generations, at least when we are concerned with operating systems, is, surprisingly, negligible.

"How will these impact the environment?" From the above we may, quite surprisingly, conclude that I/O devices having their apparent long—potential—lifespan is very positive as it reduces the need to create new devices.

"How to reuse, recycle, and dispose of these old accessories?"

In general, if a I/O device is still functional, it should still work on new operating systems, although this is dependent on new systems supporting hardware interface. I.e., it has the correctly shaped port. This being said, presently, reusing devices is quite common for mobile phones [2], although the only numbers available are for exports between nations, domestic equivalents are not known. In fact, the simple lack of such reuse statistics is why study [2] had to use international shipping to track e-waste. So, even though I/O devices should theoretically have a high degree of reuse, it is simply not known if this is truly the case. Studying this, if it even possible, is a possible area for future research.

"How to recycle these old accessories?"

Generally, most computer parts are able to be recycled, roughly 98% [31]. However, e-waste recycling is a difficult process [32] for several reasons:

- Lack of Standardized E-Waste Recycling Policies

With no clear direction on how devices must be disposed of, what devices should be recycled, and what devices are safe to simple throw away.

- Hazardous Materials

Many devices contains heavy metals that are dangerous to inhale. E.g., lead.

- Short Life-Cycles of New Devices

With relatively short lifespans, many devices are replaced by new ones, often with different materials and design. Meaning Recycling methods must try to keep up with an ever-shifting form of waste.

- Not all devices are designed to be recycled

Many devices are made with injection molded plastic (e.g., power adaptors) chassis that are not designed to be opened (e.g., AirPods Pro are glued shut [33]), or made with overly complex design; not easily separable.

The only real way to make devices more easily recyclable is enact legislation for following:

- Uniform recycling laws: Laws defining the best processes and clear standards for recycling e-waste
- Better e-waste collection
    - Better data gathering on e-waste collected
    - Better separation of e-waste into smaller classes of waste for simpler recycling
- Better standards on how devices may be built
    - That is, (reasonably) easily disassembled.

"How to dispose of these old accessories?"

Many nations, regions, and states have 'take back' laws that require different manufacturers to recycle devices that consumers bring back to them [34], however, these laws are nonuniform. That is, there is no real standard across the market for how devices should be recycled. As noted above, legislation is needed to dictate how devices on the may be made, collected, and recycled. The following figures are from the Environmental Protection Agency's monitoring of "durable goods", specifically e-waste that is generated and then collected [35].

**1960-2018 Data on Selected Consumer Electronics in MSW by Weight (in thousands of U.S. tons)**

| Management Pathway | 1960 | 1970 | 1980 | 1990 | 2000 | 2005 | 2010 | 2015 | 2017 | 2018 |
|---|---|---|---|---|---|---|---|---|---|---|
| Generation | - | - | - | - | 1,900 | 2,630 | 3,120 | 3,100 | 2,840 | 2,700 |
| Recycled | - | - | - | - | 190 | 360 | 650 | 1,230 | 1,020 | 1,040 |
| Composted | - | - | - | - | - | - | - | - | - | - |
| Combusted with Energy Recovery | - | - | - | - | - | - | - | - | - | - |
| Landfilled | - | - | - | - | - | - | - | - | - | - |

Sources: National Center for Electronics Recycling; Statistsa and Appliance Design Magazine.

*Figure 2 e-waste collected in the US*

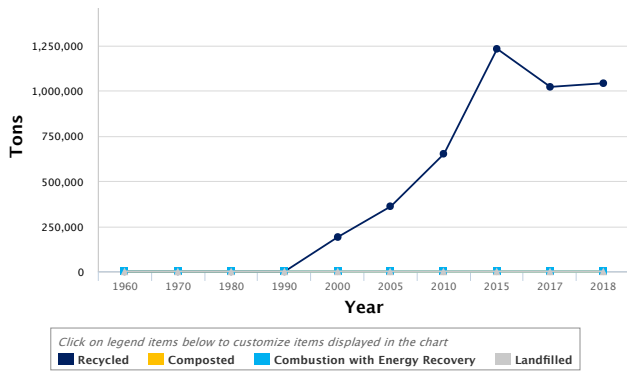

*Figure 3 Tons/year of e-waste in the US*

There is an interesting trend in *figure 2*, where despite the growing volume of generate e-waste, the volume of recycled waste is actually closing the difference does hint that recycling is seemingly catching up with production.

"What's the economic impact?"

Recycling has the potential to provide a massive amount of rare metals "40 to 800 times the amount of gold and 30 to 40 times the amount of copper mined from one metric ton of ore in the U.S" [36]. However, at present, recycling is disjointed and complex, making it difficult to be very cost effective, often being cited as "tricky" [32].

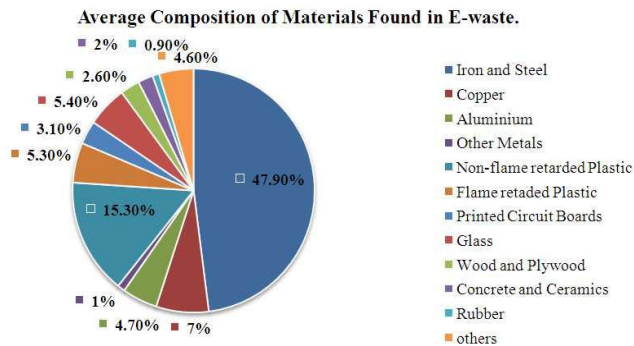

*Figure 4 Proportion of materials found in EU e-waste*

Proportion of materials from common e-waste [36]

Another thing to consider when discussing I/O devices, is that if devices are so backwards compatible, why are devices seemingly changed so often? Often Manufacturers find novel ways to make a device obsolete. This paper is more concerned with "big picture" causes of I/O obsolescence, like operating systems or connection standards, but there are other ways to generate e-waste:
- Devices with batteries that cannot be replaced by the user
- Parts that cannot be replaced or repaired by the user
- Proprietary software locking features

The proportion of devices that fall into these categories are each possible areas for future study. Of course, devices that are made obsolete does encourage spending, which in of itself is not entirely without merit, although ethically questionable in how it is encouraged. And, of course, there will be an additional cost in having old devices recycled or disposed of and an environmental cost in constructing a new device

V. Issues Encountered

There is a quite a wealth of information to draw on in this topic, although there are some areas that there is seemingly no good source on:

- Proportion of devices/cables reused domestically (in any region)
- Proportion of devices that are recycled
- Proportion of materials from recycling that actually end up in a new device
- Windows is often rather vague in what standards it may actually support (E.g., HDMI, or any other standard, is never mentioned in their documentation on displays, only DisplayPort).
- USB requires backwards compatibility for 2.0, although some devices support 1.0, although the proportion is not easily understood.
- The proportion of devices with "special" features is not easily defined, researched, nor calculated without a near exhaustive sampling of the correct market.
- The proportion of devices using nonstandard protocols is similarly not well understood.

VI. Conclusions

Major operating systems each use well defined, backwards compatible I/O standards. Devices are seemingly supported so long as a port is available for a device, and ports are becoming more and more standardized. Most notably common devices like mice, keyboards, etc. This does preclude any device that require special drivers for the operating system to use it, especially if it uses a non-standard communication protocol. E.g., not USB or Bluetooth. Power adaptors for mobile devices, like phones and laptops, in the EU will be standardized to use type c, which will help reduce e-waste dramatically. Similar legislation is needed in other regions. Similarly, legislation for recycling standards to make it more viable, as well as better statistics are needed to better understand the problem of e-waste.